\documentclass{elsart}
\usepackage{amssymb}

\begin{document}

\begin{frontmatter}
\title{Gravitational lensing time delays as a tool for testing Lorentz Invariance Violation\thanksref{grant}}

\author{Marek Biesiada},
\ead{biesiada@us.edu.pl}
\author{Aleksandra Pi\'orkowska}
\ead{apiorko@us.edu.pl}

\address{University of Silesia, Department of Astrophysics and Cosmology,\\  Uniwersytecka 4,
40-007 Katowice, Poland}

\thanks[grant]{The work was supported by the Polish State
Committee for Scientific Research Grant N N203 390034.}

\begin{abstract}

It is generally expected that quantum gravity theory will bring
the picture of a space-time foam at short distances leading to
Lorentz Invariance Violation (LIV) manifested e.g. by energy
dependent modification of standard relativistic dispersion
relation. One direction of research, pursued intensively is to
measure energy dependent time of arrival delays in photons emitted
by astrophysical sources located at cosmological distances. This
is tempered however by our ignorance of intrinsic emission delays
in different energy channels.

In this paper we discuss a test based on gravitational lensing.
Monitoring time delays between images performed in different
energy channels (e.g. optical - low energy and TeV photons) may
reveal extra delays due to distorted dispersion relation typical
in LIV theories - a test which is free from the systematics
inherent in other settings.
\end{abstract}

\end{frontmatter}

\section{Introduction}

 Despite the fact that quantum gravity theory still remains
elusive, it is generally expected that it will bring the picture
of a space-time foam at short distances leading to Lorentz
Invariance Violation (LIV) manifested e.g. by energy dependent
modification of standard relativistic dispersion relation
\cite{Amelino-Camelia, Mk 501}.

Several years ago it has been proposed to use astrophysical
objects to look for energy dependent time of arrival delays
\cite{Amelino-Camelia}. Specifically gamma ray bursts (GRBs) being
highly energetic events visible from cosmological distances are
the most promising sources of constraining LIV theories
\cite{Ellis1, Ellis2, Rodrigues,Piran Jacob}. Among other sources
the BL Lac objects like Mk 501 are considered. It is this
particular object from which 20 TeV photons were reported \cite{Mk
501}. Such objects (also called blazars) have similar nature with
quasars. In the papers quoted above some limits on LIV energy
scale have been derived (and recently corrected \cite{Piran
Errata}).

The idea of searching for time of flight delays is tempered
however by our ignorance of intrinsic delay (at source frame) in
different energy channels (see e.g. \cite{Ellis2}). Recently it
was also shown \cite{Biesiada Piorkowska}, that lack of detailed
knowledge about cosmological model (in the context of accelerating
expansion of the Universe) could be another source of systematic
effects at high redshifts.

In this paper we discuss a test based on gravitational lensing.
Its idea has been noticed in \cite{Amelino-Camelia} in the context
of gamma ray bursts playing the role of high energy photons.
Namely such a source located at a cosmological distance may
undergo gravitational lensing by a galaxy lying closer to the
observer along the line of sight with an encounter parameter small
enough to produce multiple images \cite{Falco} (the so called
strong lensing). Indeed all of known strong lensing systems
\cite{CASTLES} have quasar as a source and a galaxy (in most cases
elliptical) as a lens. The light signals emitted by a source will
be seen by the observer delayed (achromatically in classical
General Relativity) at the location of images. This opens up a
possibility to study time delays induced by LIV.
Namely, monitoring the time delays between
lensed images performed in different energy channels (e.g. optical
or gamma-ray --- low energy and TeV --- high energy photons) may
reveal extra delays due to distorted dispersion relation typical
in LIV theories. This test is free from the systematics inherent
in other settings. Next sections will substantiate the argument
further.

\section{LIV induced time delays in different cosmological models}

Following \cite{Mk 501} let us consider a phenomenological
approach for LIV  by assuming the modified dispersion relation for
photons in the form:
\begin{equation}
E^2 - p^2 c^2 = \epsilon E^2 \left( \frac{E}{{\xi}_n
E_{QG}}\right)^n \label{dispersion}
\end{equation}
where:$\epsilon = \pm 1$ is ``sign parameter'' \cite{Mk 501},
${\xi}_n$ is a dimensionless parameter. As a first guess one may
assume $E_{QG}$ equal to the Planck energy, ${\xi}_1 =1$ and
${\xi}_2=10^{-7}$ \cite{Piran Jacob}. The dispersion relation
(\ref{dispersion}) essentially corresponds to the power-law
expansion (see \cite{Ellis1}) so for practical purposes (due to
smallness of expansion parameter $E/E_{QG}$) only the lowest terms
of the expansion are relevant. Because in some LIV theories the
odd power terms might be forbidden \cite{Burgess} usually the
cases of $n=1$ and $n=2$ are retained. It should be noted that in
some string theories third order corrections appear as leading
ones.

The relation (\ref{dispersion}) leads to a hamiltonian
\begin{equation}
{\cal H} = \sqrt{p^2 c^2  \lbrack 1 + \epsilon \left(
\frac{E}{{\xi}_n E_{QG}} \right) ^n \rbrack } \label{hamiltonian}
\end{equation}
from which time dependent group velocity $v(t) = \frac{\partial
{\cal H}}{\partial p}$ can be inferred.

The comoving distance travelled by photon to the Earth is
\begin{equation}
r(t) = \int_{t_{emission}}^{t_{detection}} v(t) dt = \int_{0}^{z} v(z')
\frac{dz'}{H(z') (1+z')} \label{distance1}
\end{equation}
where in the last equation a standard time-redshift
parametrization was taken into account. $H(z)$ here denotes the
Universe's expansion rate (so called Hubble function). Starting
from this point our considerations will have cosmological
connotation. The reason for this is simple --- the modifications
due to LIV theories are really tiny, so one has to look for
sources located at cosmological distances (such like quasars or
gamma ray bursts) which are far enough to compensate for the
smallness of LIV corrections. This means that cosmological
background geometry should be taken into account. From now on we
will assume it to be a flat Friedmann-Robertson-Walker model with
$\Omega_m=0.3$ and $\Omega_{\Lambda} = 0.7$ --- the so called
concordance model, as supported by observations \cite{Spergel}.
Some alternatives to the concordance model could in principle be
discussed as well \cite{Biesiada Piorkowska}.

Expressing group velocity in terms of redshift, we get
\begin{equation}
v(z) \simeq c(1+z) \lbrack 1 + \epsilon \frac{(n+1)}{2}  \left(
\frac{E}{\xi_{n} E_{QG}} \right)^n (1+z)^n \rbrack
\label{velocity2}
\end{equation}
Time of flight for the photon of energy $E$ is equal to
\begin{equation}
t_{LIV} = \int_{0}^{z} \lbrack 1 + \epsilon \frac{n+1}{2} \left(
\frac{E}{{\xi}_n E_{QG}} \right)^n (1+z')^n \rbrack
\frac{dz'}{H(z')} \label{flight time}
\end{equation}
In the first term one easily recognizes the time of flight for
photons in standard relativistic cosmology (i.e. without LIV). Due
to very small magnitude of LIV corrections it also fairly
represents the time of flight for low energy photons. Therefore
below we neglect LIV corrections at low energy.

Consequently, the time delay between a low energy and a high
energy photon is equal to
\begin{equation}
\Delta t_{LIV} =   \frac{n+1}{2} \left( \frac{E}{{\xi}_n E_{QG}}
 \right)^n  \int_0^z  \frac{(1+z')^n dz'}{H(z')} \label{time delay}
\end{equation}
where we restricted our attention to ``infraluminal'' motion of
high energy photons (i.e. low energy photons arrive earlier to the
observer). Generalization to ``superluminal'' motion is
straightforward --- the same value with just an opposite sign
(time delays become early arrivals).

The idea of observational strategy emerging form (\ref{time
delay}) is again simple: monitor appropriate (i.e. emitting both
low and high energy photons) cosmological source at different
energy channels and try to detect this time delay. Some attempts
along this line have already been undertaken \cite{Ellis1, Ellis2,
Ellis new}. However there remains an indispensable uncertainty
about intrinsic time delays: there is no reason for which low and
high energy signal should be emitted simultaneously, and while
detecting distinct signals (peaks in the light curve) at different
energies we have no idea which one was sent first.

Our method outlined below, invoking gravitational lensing allows
to get rid of this ambiguity. Before describing it in subsequent
paragraph let us recall now that in cosmology one distinguishes
three types of distances:

(i) comoving distance:
\begin{equation} \label{comoving}
r(z) = c \int_0^z \frac{dz'}{H(z')} = \frac{c}{H_0} \tilde{r}(z)
\end{equation}
by $\tilde{r}(z)$ we denoted a reduced (dimensionless) comoving
distance, i.e. a comoving distance expressed as a fraction of the
Hubble horizon $d_H = c/H_0$,

 (ii) angular diameter distance:
\begin{equation} \label{angular}
D_A(z) = \frac{1}{1+z} r(z)
\end{equation}

(iii) luminosity distance:
\begin{equation} \label{luminosity}
D_L(z) = (1+z) r(z)
\end{equation}
Angular diameter distance is the one used in gravitational lensing
theory (because gravitational lensing deals with light deflection
i.e. essentially with angles). The luminosity distance is a
measure invoked while using standard candles (e.g. SNIa). The
point is that both distance measures are related to the comoving
distance by a $1+z$ factor (see above). The comoving distance,
then, is closely related to the time of flight $r(z) = c t.$ It is
in fact the distance to the source measured in light years.
Therefore we can rewrite the time of flight in LIV theory
(\ref{flight time}) in terms of comoving distance: $r_{LIV}(z) = c
t_{LIV}$. Of course the comoving distance to the source is fixed
--- there are photons of different energies that travel with
different speed. It is however useful (for later calculations) to
think as if they travelled with the same speed $c$ but along
different comoving distances $r(z)$ and $r_{LIV}(z)$.

\section{Gravitational lensing time delays}

Gravitational lensing of quasars and extragalactic radio sources
at high redshifts by foreground galaxies is now well established
and has developed into a mature branch of both theoretical and
observational astrophysics \cite{Falco}. Misalignment of the
source, the lens and observer results typically in multiple images
whose angular positions and magnification ratios allow
reconstructing lensing mass distribution. In particular they
provide an independent confirmation of dark matter in galaxies and
became an important tool for investigating dark matter
distribution. Another important ingredient of gravitational
lensing is the time delay between lensed images of the source.
This effect originates as a competition between Shapiro time delay
from the gravitational field and the geometric delay due to
bending the light rays and is best understood in terms of Fermat
principle. In other words, the intervening mass between the source
and the observer introduces an effective index of refraction,
thereby increasing the light travel time.

In general, the light travel time can be calculated as
\begin{equation} \label{travel time}
t(\overrightarrow{x}) = \frac{1+z_l}{c} \frac{D_l
D_s}{D_{ls}}\left[ \frac{1}{2}
\left(\overrightarrow{x}-\overrightarrow{{\beta}}\right)^2 -
\psi(\overrightarrow{x})\right]
\end{equation}
where: $\overrightarrow{x}$ and $\overrightarrow{\beta}$ are
positions (as projected on the celestial sphere) of the image and
the source, $\psi(\overrightarrow{x})$ is the projected
gravitational potential (i.e. the actual potential integrated
along line of sight), $D_l$, $D_s$ are angular diameter distances
to the lens and the source located at redshifts $z_l$ and $z_s$
respectively ($D_{ls}$ is the angular diameter distance between
lens and source). From now on we adopt the notation (standard in
gravitational lensing theory) where $D$ denotes angular diameter
distance, and subscripts refer to the components of lensing system
(i.e. the source, the lens or the observer).

The lensing is called strong if source position happens to lie
within the so called Einstein ring --- the circle of a radius
$\vartheta_E$ (defining the proper deflection scale of a given
lens). In this case multiple images appear and since lensing
galaxies are often ellipticals, the number of images is usually
equal to four \cite{CASTLES, Ratnatunga95} (or five \cite{fifth}
-- the issue of image multiplicity is discussed e.g. in
\cite{Falco}). However, the surprisingly realistic model of the
lens potential is that of a singular isothermal sphere (SIS)
\cite{Falco}. Indeed lensing by elipticals can be modelled by its
variant called singular isothermal ellipsoid (SIE). Therefore for
the purpose of illustrating our ideas we shall restrict our
attention to the SIS model since generalization to SIE is rather
straightforward and would not change our conclusions.

The Einstein ring radius for the SIS model is:
\begin{equation} \label{Einstein SIS}
\vartheta_E =  4 \pi \frac{D_{ls}}{D_s} \frac{\sigma^2}{c^2}
\end{equation}
where $\sigma$ denotes one-dimensional velocity dispersion of
stars in lensing galaxy. If the lensing is strong i.e.
$\beta:=|\overrightarrow{\beta}|< \vartheta_E$ then two co-linear
images A and B form on the opposite side of the lens, at radial
distances $R_A = \beta + \vartheta_E$ and $ R_B = \vartheta_E -
\beta$ having time delays between the images:

\begin{equation} \label{SIS delay}
\Delta t_{SIS} = \frac{1+z_l}{ 2 c} \frac{D_l D_s}{D_{ls}}(R_A^2 -
R_B^2)
\end{equation}
which according to the above mentioned relations for SIS model can
also be written as
\begin{equation} \label{SIS other way}
\Delta t_{SIS} = \frac{2(1+z_l)}{c}\frac{D_l D_s}{D_{ls}}
\vartheta_E \beta
 = \frac{8 \pi}{H_0} \widetilde{r}_l \beta
\frac{\sigma^2}{c^2}
\end{equation}
In the last equation $\tilde{r}_l$ denotes  the reduced comoving
distance to the lens. The equation (\ref{SIS delay}) is commonly
used by gravitational lensing community because it reduces time
delay problem to relative astrometry of images, whereas $\beta$ is
much harder to asses (it is small in order for strong lensing to
occur) and Einstein ring radius is not a directly observable
quantity. However, the equation (\ref{SIS other way}) is more
useful from the theoretical point of view. In particular it shows
explicitly that the time delay between images is created at the
lens location ($\widetilde{r}_l$ factor). Let us stress again that
this time delay is achromatic in General Relativity.

\section{LIV induced time delays and gravitational lensing time delays}

Let us now imagine a source at cosmological distance emitting low
energy and high energy (in TeV range) photons which undergoes
gravitational lensing by a foreground galaxy. Let us also assume
that LIV type distorted dispersion relation (\ref{dispersion})
holds in nature. The observer would notice again time delays
between images, but this time it would be a combined effect of
gravitational lensing and LIV. Therefore it would no longer be
achromatic. This idea was formulated originally in
\cite{Amelino-Camelia} but up to our best knowledge it has not
been further developed.

It is rather straightforward to calculate this by using the above
mentioned fictitious ``LIV comoving distance'' $r_{LIV}(z)$,
namely:
\begin{equation} \label{LIV SIS delay}
\Delta t_{LIV,SIS} =  \frac{8 \pi}{H_0} \widetilde{r}_{LIV}(z_l)
\beta \frac{\sigma^2}{c^2}
\end{equation}
where:
\begin{equation} \label{r LIV}
\widetilde{r}_{LIV}(z_l) = \widetilde{r}_l + H_0 \frac{n+1}{2}
\left( \frac{E}{{\xi}_n E_{QG}}
 \right)^n  \int_0^{z_l}  \frac{(1+z')^n dz'}{H(z')}
\end{equation}
Because the LIV effect is extremely small, let us restrict further
to the $n=1$ case:
\begin{equation} \label{r LIV1}
\widetilde{r}_{LIV}(z_l) = \widetilde{r}_l + H_0 \frac{E}{E_{QG}}
 \int_0^{z_l}  \frac{(1+z') dz'}{H(z')}
\end{equation}

Now we can assume that observations in low energy would
essentially provide time delay between images equal to $\Delta
t_{SIS}$, whereas monitoring of the same images in high energy
(TeV) channel would provide $\Delta t_{LIV,SIS}$. These two
measurements would differ by
\begin{equation} \label{LIV effect}
\Delta t_{LIV,SIS} - \Delta t_{SIS}=  \frac{8 \pi}{H_0} \beta
\frac{\sigma^2}{c^2} \frac{E}{E_{QG}}
 \int_0^z  \frac{(1+z') dz'}{H(z')}
\end{equation}

Rigorously one should rather calculate $\Delta t_{LIV,SIS}(E_1) -
\Delta t_{LIV,SIS}(E_2)$ and write $\Delta E = E_1 - E_2$ in the
numerator of (\ref{LIV effect}) but due to our assumption that
$E_2$ is many orders of magnitude smaller than $E_1$ and thus
having negligible LIV correction the above expression is a good
appproximation.
 Let us make an estimate for the above LIV effect
taking a real strong lensing system. HST 14176+5226 can serve as
an example. This system was discovered with the Hubble Space
Telescope \cite{Ratnatunga95} and further confirmed to be a
gravitational lens \cite{Crampton}. The lensed source is a quasar
at redshift $z_s = 3.4$ whereas the lens is an elliptical galaxy
having redshift $z_l = 0.809$. The lens model best fitted to the
observed images was based on a singular isothermal ellipsoid
\cite{Ratnatunga99} giving the Einstein radius $\theta_E =
1''.489$ and $\beta = 0".13 = 8.4 \times 10^{-7} \;rad$.

Optical spectroscopy of the lensing galaxy in HST 14176+5226
system \cite{Subaru} provided measurements of the velocity
dispersion in lensing galaxy. These measurements have been
confirmed by Treu and Koopmans \cite{Treu} who performed
spectroscopic observations on Keck telescope as part of Lenses
Structure and Dynamics (LSD) Survey.  The result is $\sigma = 290
\pm 8 \; km/s$.

Substituting these data to (\ref{LIV effect}) gives $\Delta
t_{LIV,SIS} - \Delta t_{SIS}$ equal to $3.7 \times 10^{-9} \;s$
for $5 \; TeV$ photons and $1.5 \times 10^{-8} \; s$ for $20 \;
TeV$ ones.

The model presented above was the simplest one just because its
aim was to illustrate ideas. So was also the purpose of choosing
the HST 14176+5226 system as an example. In reality one would
encounter systems with different numbers of images or different
separations, e.g. SDSS J1004+4112 \cite{ls1} with an image
separation of $14".6$ or recently discovered system SDSS
J1029+2623, where a quasar at $z_s = 2.197$ is doubly imaged by a
massive galaxy cluster at $z_l = 0.55$ with images separation of
$22".5$ \cite{ls2}.

However none of these effects is crucial to our general arguments
because of their differential setting ($\Delta t_{high\; energy}$
vs. $\Delta t_{low\; energy}$ ). High and low energy photons from
each image travel along the same paths (respectively) thus
suffering the same shear effects (respectively). We do not mean
here that one could disregard these effects but rather to
strengthen the standard lore of lensing community to treat each
lensing system separately in a detailed manner.

Closing this section it would be interesting to ask how the LIV
effects might modify image configurations. It could be suspected
that they might do so since from the Fermat's principle
perspective images are located at stationary points of the
wavefront travel time functional (given by equation (\ref{travel
time})). Therefore since LIV modifies time of flight in an energy
dependent way (due to modified dispersion relation) then one
expects the images seen at different energies located at different
positions. It is easy to see that for the SIS lens
(generalizations to other mass profiles are also rather
straightforward) the difference between Einstein radii for high
and low energy photons $\Delta \theta_{E,LIV} := \theta_{E,LIV} -
\theta_E$ would be given by formula:
\begin{equation} \label{Einstein radius difference}
\Delta \theta_{E,LIV} = \theta_E \; \frac{E}{E_{QG}} \left(
\frac{I^{(1)}(z_l,z_s)}{\widetilde{r}(z_l,z_s)} -
\frac{I^{(1)}(z_s)}{\widetilde{r}(z_s)}\right)
\end{equation}
where: $I^{(1)}(z_1,z_2) :=  \int_{z_1}^{z_2}  \frac{(1+z')
dz'}{H(z')}$. For realistic lens configurations like HST
14176+5226 this would give negligibly small corrections of order
$10^{-16} \; arc\;sec$. Hence even if LIV were operating this
would not be able to change macro-images position in a detectable
way. However it cannot be excluded that such minute differences
could become relevant while studying caustic crossing possibly
leading to different magnification patterns due to microlensing at
different energies.

\section{Discussion and conclusions}

In this paper we discussed a method (first noticed in
\cite{Amelino-Camelia}) to test LIV effects by monitoring time
delays between images of gravitationally lensed quasars in low and
high energy channels. In standard theory (General Relativity) the
result should be the same
--- gravitational lensing is essentially achromatic. On the other
hand in the presence of LIV effects time delays loose this
property --- high energy photons should come at different times
comparing with low energy ones. Therefore time delays between
images should be different at different energies (e.g. optical or
gamma-rays and TeV photons). We are tacitly assuming (following
approach taken by the rest of LIV studying community) that LIV
effects are manifested only in high energy domain (where the small
scale ``foamy'' structure of the space-time reveals itself)
whereas the overall background geometry of space-time shaped by
low energy content of the Universe is that of General Relativity
(more precisely
--- the flat Friedman-Robertson-Walker model as suggested by
cosmological data) with light deflection (i.e. geodesic motions)
defined in a standard way.

Because this method is differential in nature, it gets rid of the
assumptions about intrinsic time delays of signals at different
energies. In fact time delays between images at different energies
could be established in different experiments (at unrelated
observing sessions) performed on given lensing system. The only
demand is that they are accurate enough (done with a sufficient
temporal resolution). Since the time delay between images is
produced at the lens location, the result does not depend very
strongly on the cosmological model. Lenses are located at modest
redshifts where all realistic cosmological models essentially
agree.

One may ask if appropriate lensing systems (i.e. having sources
emitting both low and high energy photons) exist. It is an
observational fact that very high energy emission ($E > 100 \;
GeV$) has been detected from over a dozen of blazars \cite{Wagner
2007} which have similar nature with quasars. Quasars, on the
other hand are the sources in all known strong lensing systems ---
CASTLES database contains a 100 of such systems \cite{CASTLES}. It
is a matter of coordinating strong lensing surveys with
experiments in high energy astrophysics (such like AGILE, GLAST or
MAGIC experiments \cite{blazar}) and the future will certainly
bring the discovery of lensed high energy source. Angular
resolution of high energy experiments is gradually being improved.
For example a recently launched AGILE instrument \cite{Tavani} has
been designed to obtain accurate localization ($\sim 2' - 3'$ ) of
transient events by the Gamma Ray Imaging Detector (GRID) - SA
combination. So it is already close to typical image separation
and definitely being able to see images like those from SDSS
J1004+4112 or SDSS J1029+2623 as separated. Depending on exposure
and the diffuse background its flux sensitivity threshold can
reach values of $(10-20)\times 10^{-8} \; photons \;cm^{-2}\;
s^{-1}$ at energies higher than $100\; MeV$ with with an effective
area above $200\; cm^2$ at $30\; MeV$. Moreover it has excellent
timing capability, with overall photon absolute time tagging of
uncertainty below $2\; \mu s$ and very small deadtimes ($< 200\;
\mu s$ for the GRID,  $\sim 5\; \mu s$ for the sum of the SA
readout units, and $\sim 20\; \mu s$ for each of the individual
CsI bars). In fact, AGILE instrument is optimized in the range
below $1\; GeV$ hence it is not representative to $TeV$ range
experiments needed to probe LIV but it clearly shows the the
gradual improvement of sensitivity, timing and angular resolution
in high energy astrophysics.

An order of magnitude estimate for the effect discussed in this
paper is not encouraging now. For a typical lensing system like
HST 14176+5226 it is of order of nano-seconds. However, having in
mind that high energy astrophysical sources display rapid
variability (indeed intrinsic variability in relativistic shocks
powering these sources is enhanced by a Lorentz factor typically
of order of $10^2$) and that e.g. light curves of gamma-ray bursts
are already sampled with mili-second resolution (and AGILE went
down to micro-seconds) one should not reject the idea presented
above on the grounds that it is not within the scope of
present-day observational technology.

\end{document}